\documentclass[longauth,traditabstract]{aa}

\def\del#1{}  \def\remark#1{}

\def\symbolfootnote[#1]#2{\begingroup\def\thefootnote{\fnsymbol{footnote}}\footnote[#1]{#2}\endgroup} 
\newcommand{\rmn}{\mathrm}
\newcommand{\CR}{\mathrm{CR}}
\newcommand{\expval}[1]{\left\langle #1 \right\rangle}

\newcommand{\dps}{\displaystyle}
\newcommand{\eps}{\varepsilon}
\usepackage{graphicx}
\usepackage{natbib}
\usepackage{txfonts}
\begin{document}

\title{Constraining Cosmic Rays and Magnetic Fields in the Perseus Galaxy
  Cluster with TeV observations by the MAGIC telescopes}

\author{
 J.~Aleksi\'c\inst{1} \and
 E.~A.~Alvarez\inst{2} \and
 L.~A.~Antonelli\inst{3} \and
 P.~Antoranz\inst{4} \and
 M.~Asensio\inst{2} \and
 M.~Backes\inst{5} \and
 U.~Barres de Almeida\inst{6} \and
 J.~A.~Barrio\inst{2} \and
 D.~Bastieri\inst{7} \and
 J.~Becerra Gonz\'alez\inst{8,}\inst{9}
% ...
}
\institute { IFAE, Edifici Cn., Campus UAB, E-08193 Bellaterra, Spain
 \and Universidad Complutense, E-28040 Madrid, Spain
 \and INAF National Institute for Astrophysics, I-00136 Rome, Italy
 \and Universit\`a  di Siena, and INFN Pisa, I-53100 Siena, Italy
 \and Technische Universit\"at Dortmund, D-44221 Dortmund, Germany
 \and Max-Planck-Institut f\"ur Physik, D-80805 M\"unchen, Germany
 \and Universit\`a di Padova and INFN, I-35131 Padova, Italy
 \and Inst. de Astrof\'{\i}sica de Canarias, E-38200 La Laguna, Tenerife, Spain
 \and Depto. de Astrof\'{\i}sica, Universidad de La Laguna, E-38206 La Laguna, Spain
% ...
}

\date{Received 22/10/2011 / Accepted 06/03/2012}


\abstract{Galaxy clusters are being assembled today in the most energetic phase
  of hierarchical structure formation which manifests itself in powerful shocks
  that contribute to a substantial energy density of cosmic rays (CRs). Hence,
  clusters are expected to be luminous gamma-ray emitters since they also act as
  energy reservoirs for additional CR sources, such as active galactic nuclei
  and supernova-driven galactic winds. To detect the gamma-ray emission from CR
  interactions with the ambient cluster gas, we conducted the deepest to date
  observational campaign targeting a galaxy cluster at very high-energy
  gamma-rays and observed the Perseus cluster with the MAGIC Cherenkov
  telescopes for a total of $\sim85$\,h of effective observing time.  This
  campaign resulted in the detection of the central radio galaxy NGC~1275 at
  energies $E> 100$~GeV with a very steep energy spectrum.  Here, we restrict
  our analysis to energies $E> 630$~GeV and detect no significant gamma-ray
  excess.  This constrains the average CR-to-thermal pressure ratio to be
  $\lesssim1$--2\%, depending on assumptions and the model for CR
  emission. Comparing these gamma-ray upper limits to models inferred from
  cosmological cluster simulations that include CRs constrains the maximum CR
  acceleration efficiency at structure formation shocks to be $<50\%$.
  Alternatively, this may argue for non-negligible CR transport processes such
  as CR streaming and diffusion into the outer cluster regions.  Finally, we
  derive lower limits on the magnetic field distribution assuming that the
  Perseus radio mini-halo is generated by secondary electrons/positrons that are
  created in hadronic CR interactions: assuming a spectrum of $E^{-2.2}$ around
  TeV energies as implied by cluster simulations, we limit the central magnetic
  field to be $>$ 4--9 $\mu$G, depending on the rate of decline of the magnetic
  field strength toward larger radii. This range is well below field strengths
  inferred from Faraday rotation measurements in cool cores. Hence, the hadronic
  model remains a plausible explanation of the Perseus radio mini-halo.}

\keywords{Gamma rays: galaxies: clusters: individual: Perseus} 

\titlerunning{Constraining Cosmic Rays in the Perseus Cluster with MAGIC}

\authorrunning{Aleksi\'c et al.}

\maketitle

\section{Introduction}

In the hierarchical model of structure formation, clusters of galaxies are
presently forming through mergers of smaller galaxy groups, and assembling
masses up to a few times $10^{15} \rmn{M}_\odot$. During these merger events,
enormous amounts of energy --- of the order of the final gas binding energy
$E_\rmn{bind} \sim 3 \times (10^{61}$--$10^{63})$~erg --- are released through
collisionless shocks and turbulence. This energy is dissipated
\begingroup
\let\thefootnote\relax\footnotetext{* Corresponding authors: F. Zandanel (fabio@iaa.es), C. Pfrommer (christoph.pfrommer@h-its.org), P. Colin (colin@mppmu.mpg.de), A. Pinzke (apinzke@physics.ucsb.edu) \& S. Lombardi (saverio.lombardi@pd.infn.it)}
\endgroup
on a dynamical timescale of $\tau_\rmn{dyn} \sim 1$~Gyr \citep[see][for a
review]{2005RvMP...77..207V}.  Hence, the corresponding energy dissipation rates
are $L \sim (10^{45}$--$10^{47})\mbox{ erg s}^{-1}$. If only a small fraction of
this energy were converted into non-thermal particle populations, the associated
emission should be detectable in the gamma-ray regime and can potentially be
used to decipher the short- and long-term history of structure formation.

Many galaxy clusters show large scale diffuse synchrotron radio emission in the
form of so-called radio (mini-)halos which proves the existence of magnetic
fields and relativistic electrons permeating the intra-cluster medium (ICM)
\citep[e.g.,][]{2004rcfg.procE..25K, 2004NewAR..48.1137F}. Through this
synchrotron emission we can identify the sites of acceleration and injection of
the relativistic electrons into the ICM: (1) merger shocks propagating toward
the cluster outskirts, observed as giant radio relics, (2) active galactic
nuclei (AGN) in clusters, observed through radio jets and bubbles, and (3)
galactic winds and ram pressure stripping are often accompanied by synchrotron
emission. In analogy with shocks within our Galaxy, such as those in supernova
remnants, galaxy clusters should also be acceleration sites for relativistic
protons and heavier relativistic nuclei.  Due to their higher masses compared
with the electrons, protons and nuclei are accelerated more efficiently to
relativistic energies and are expected to show a ratio of the spectral energy
flux of cosmic ray (CR) protons to CR electrons above 1 GeV of about 100
\citep[as it is observed in our Galaxy between 1--10 GeV,
see][]{2002cra..book.....S}. CR protons also have radiative cooling times that
are larger than the corresponding cooling times of CR electrons by the square of
the mass ratio, $(m_\rmn{p}/m_\rmn{e})^2$, and hence can accumulate for the
Hubble time in a galaxy cluster \citep{1996SSRv...75..279V}. For gas densities
in galaxy clusters, $n$, the radiative cooling time of CR protons is much longer
relative to the hadronic timescale, $\tau_\rmn{pp}\simeq 30 \,\rmn{Gyr}\times
(n/10^{-3}\,\rmn{cm}^{-3})^{-1}$, on which CR protons collide inelastically with
ambient gas protons of the ICM. This is the process in which we are primarily
interested here as it produces pions that successively decay into synchrotron
emitting electrons/positrons (so-called secondaries) and gamma-rays.

There are two principal models that explain radio \mbox{(mini-)}halos.  In the
``hadronic model'' the radio emitting electrons are produced in inelastic CR
proton-proton (p-p) interactions \citep{1980ApJ...239L..93D,
  1982AJ.....87.1266V, 1997ApJ...477..560E, 1999APh....12..169B,
  2000A&A...362..151D, 2001ApJ...559...59M, 2003MNRAS.342.1009M,
  2003A&A...407L..73P, 2004A&A...413...17P, 2004MNRAS.352...76P,
  2007IJMPA..22..681B, 2008MNRAS.385.1211P, 2008MNRAS.385.1242P,
  2009JCAP...09..024K, 2010MNRAS.401...47D, 2010arXiv1003.0336D,
  2010arXiv1003.1133K, 2010arXiv1011.0729K, 2011A&A...527A..99E,2012ApJ...746...53F} 
requiring only a very modest fraction of a few percent of CR-to-thermal pressure. Some
clusters, known as cool-core clusters, e.g., the \object{Perseus cluster} studied here,
are characterized by a central temperature decrease and a correspondingly
enhanced central ICM density \citep{1994ARA&A..32..277F,2010A&A...513A..37H}.
These cool cores provide high target densities for hadronic CR interactions,
increasing the resulting gamma-ray flux.  This is the main reason for a larger
expected flux from Perseus in comparison to other nearby clusters
\citep{2010MNRAS.409..449P, 2011PhRvD..84l3509P}.  Another alternative is the
``re-acceleration model'': during states of powerful ICM turbulence, e.g., after
a cluster merger, re-acceleration might be efficient enough to accelerate CR
electrons to high enough energies ($\sim10$~GeV) that they produce observable
radio emission \citep{1987A&A...182...21S, 1993ApJ...406..399G,
  2002A&A...386..456G, 2005MNRAS.363.1173B, 2007MNRAS.378..245B,
  2010arXiv1008.0184B, 2009A&A...507..661B}. This, however, requires a
sufficiently long-lived CR ICM electron population at energies around 100 MeV
which might be maintained by re-acceleration processes by particle-plasma wave
interactions at a rate faster than the cooling processes.  We refer the reader
to \citet{2011A&A...527A..99E} for a discussion on the strengths and weaknesses
of each of the models.

That magnetic fields with central field strengths ranging between $B \sim
(1-20)$~$\mu$G are ubiquitous in galaxy clusters and permeate the ICM is
suggested by Faraday rotation measurements \citep[RM;][]{1991ApJ...379...80K,
  2001ApJ...547L.111C, 2002ARA&A..40..319C,2005A&A...434...67V,
  2011A&A...529A..13K}. Combining the diffuse synchrotron radio emission with
detections of galaxy clusters in hard X-rays can also be used to constrain the
intra-cluster magnetic field if the hard X-ray emission is due to inverse
Compton up-scattering of cosmic microwave photons by CR electrons (see
\citealp{2008SSRv..134...71R} for a recent review). The derived magnetic field
strengths, typically 0.1 $\mu$G, are in direct conflict with the much stronger
field strengths estimated from Faraday RM analyses.  However, more sensitive
observations by Swift/BAT have not found evidence of a hard power-law tail above
the thermal emission. With the potential exception of the Bullet Cluster (i.e.,
1E 0657-558), the emission is compatible with a multi-temperature plasma
\citep{2009ApJ...690..367A, 2010ApJ...725.1688A, 2011ApJ...727..119W}, disputing
earlier claims of the existence of a nonthermal component. Combining
hadronically-induced synchrotron emission by secondaries from CR interactions
with the associated pion-decay gamma-ray emission would yield significant
information on cluster magnetic fields \citep[e.g.,][]{2004MNRAS.352...76P,
  2008MNRAS.385.1242P, 2011ApJ...728...53J}. Limits on the gamma-ray emission
provide limits on the intra-cluster magnetic field if the hadronic model is
applicable to the radio \mbox{(mini-)}halo emission.

Very high-energy (VHE, $E>100$~GeV) gamma-ray observations are very important
for delimiting the CR pressure. A substantial CR pressure contribution can bias
hydrostatic equilibrium cluster masses as well as the total Comptonization
parameter due to the Sunyaev-Zel'dovich effect which is proportional to the
thermal energy of a galaxy cluster (e.g., see
\citealp{2007MNRAS.378..385P}). This would at least complicate, if not render it
impossible, to use these methods for estimating cosmological parameters with
galaxy clusters.  There are two ways to assess the bias due to nonthermal
pressure contributions from CRs, magnetic fields or turbulence.  At the cluster
center a comparison of X-ray and optically-inferred gravitational potential
profiles yields an upper limit of 0.2-0.3 for the ratio of nonthermal pressure
to the thermal gas pressure \citep{Churazov_etal:2008, Churazov_etal:2010}. On
large scales, there are indications that hydrostatic mass estimates can be
biased by up to $\sim20\%$ relative to weak gravitational lensing masses at
$R_{500}$ --- the radius of a sphere enclosing a mean density that is 500 times
the critical density of the Universe \citep{Mahdavi_etal:2008}. While this could
be due to nonthermal pressure, weak lensing mass estimates must be interpreted
with caution since they are intrinsically biased: cluster mass (shear) profiles
at large radii attain systematic modifications due to substructure and halo
triaxiality \citep{2011ApJ...740...25B}.  VHE gamma-ray emission is a
complementary method for constraining the pressure contribution of CRs that is
most sensitive to the cluster core region. Multi-frequency constraints limited
  the CR pressure component to be below a few percent of the thermal pressure in
  the best cases \citep[e.g.,][]{2004A&A...413...17P, 2004A&A...424..773R,
    2010ApJ...717L..71A, 2011PhRvD..84l3509P}.  However, this approach assumes
that the CR component is fully mixed with the ICM and is almost insensitive to a
two-phase structure of CRs and the thermal ICM.  In addition to this
cosmological motivation, constraining the CR pressure contribution in clusters
in combination with cosmological hydrodynamical simulations provides limits on
the average acceleration efficiency at strong formation shocks, the CR transport
properties \citep{2011A&A...527A..99E} and the impact of the cluster's dynamical
state on the CR population.

Following up on earlier high-energy gamma-ray observations of galaxy clusters,
in this work we present the deepest yet VHE~gamma-ray observation of a galaxy
cluster (for space-based cluster observations in the GeV-band, see
\citealt{2003ApJ...588..155R, 2010ApJ...717L..71A, 2010JCAP...05..025A}; for
ground-based observations in the energy band $>100$~GeV, see
\citealt{2006ApJ...644..148P, 2008AIPC.1085..569P, 2009A&A...495...27A,
  2009arXiv0907.0727T, 2009arXiv0907.3001D, 2009arXiv0907.5000G,
  cangaroo_clusters,2009ApJ...706L.275A,2010ApJ...710..634A}).  We are able to
constrain simulation-based models for the CR induced gamma-ray emission and the
cluster magnetic field in the hadronic model for radio (mini-)halos.

The MAGIC Telescopes observed the Perseus cluster in stereoscopic mode from
October 2009 to February 2011 for a total of $\sim85$\,hr of effective
observation time. This campaign produced the VHE detection of the head-tail
radio galaxy \object{IC~310}, which lies in the field of view of the Perseus observation
\citep{2010ApJ...723L.207A}, and of the central radio galaxy \object{NGC~1275}
\citep{2011arXiv1112.3917M}.  
Here, we focus on the CR induced
gamma-ray emission from the cluster itself. Galaxy clusters are also very
promising targets for constraining the dark matter annihilation cross section or
decay rate \citep{2006A&A...455...21C, 2009arXiv0905.1948P,
    2008arXiv0812.0597J, 2010arXiv1007.3469C,
    2010arXiv1009.5988D,2011arXiv1104.3530S,
    2011PhRvD..84l3509P,2011arXiv1107.1916G}.  We defer any such analysis to
future work. The presence of a central gamma-ray emitting source at energies
$<600$~GeV, combined with the expected flat dark matter annihilation emission
profile out to the virial radius owing to the substructures that dominate the
cluster emission
\citep{2011arXiv1104.3530S,2011PhRvD..84l3509P,2011arXiv1107.1916G}, calls for
novel analysis techniques that are beyond the scope of the present study.

This work has two companion papers. One is dedicated to the NGC~1275 detection
using the August 2010 - February 2011 campaign data \citep{2011arXiv1112.3917M},
while the other will address the NGC~1275 multi-wavelength emission from radio
to VHE and give a comprehensive interpretations of all collected data (MAGIC Coll., 
in prep.). The paper is organized as follows. In
Sect.~\ref{sec:Coma-Perseus}, we explain and present our target selection and
put the observation of the cool-core cluster Perseus into the broader context of
nonthermal cluster radio emission.  Specifically, we compare Perseus to the
expected TeV characteristics of a merging cluster, namely Coma.  In
Sect.~\ref{sec:obs}, we describe the MAGIC telescopes, present the observations,
and detail the data analysis and results. These data are used to constrain the
CR population in the Perseus cluster and the magnetic field distribution in the
hadronic model of the radio mini-halo (Sect.~\ref{sec:CR}).  Finally, in
Sect.~\ref{sec:conclusions}, we summarize our findings.  All quantities are
scaled to the currently favored value of Hubble's constant $H_{0} =
70$~km~s$^{-1}$~Mpc$^{-1}$.

\section{Target selection: comparing the nonthermal  emission of
  cool-core versus merging clusters}
\label{sec:Coma-Perseus}

The Perseus cluster was selected for the MAGIC observations as it is the most
promising target for the detection of gamma-rays coming from neutral pion decay
resulting from hadronic CR interactions with the ICM \citep{2010ApJ...710..634A,
  2010MNRAS.409..449P,2011PhRvD..84l3509P}. This cluster of galaxies (also known
as Abell~426), at a distance of 77.7\,Mpc ($z =0.018$), is the brightest X-ray
cluster with a luminosity in the soft X-ray band (ranging from $0.1-2.4$~keV),
$\mathrm{L_{X,0.1-2.4}} = 8.3\times10^{44}$\,erg\,s$^{-1}$
\citep{2002ApJ...567..716R}. It contains a massive cool core with high central
gas densities of about $0.05$\,cm$^{-3}$ \citep{2003ApJ...590..225C}. The
cluster also contains a luminous radio mini-halo with an extension of $200$~kpc
\citep{1990MNRAS.246..477P}. More details on possible VHE~gamma-ray emission in
galaxy clusters can be found in the first MAGIC paper about the Perseus cluster
observation with a single telescope \citep{2010ApJ...710..634A}.

\begin{figure*}[hbt!]
\centering
\includegraphics[width=0.48\textwidth]{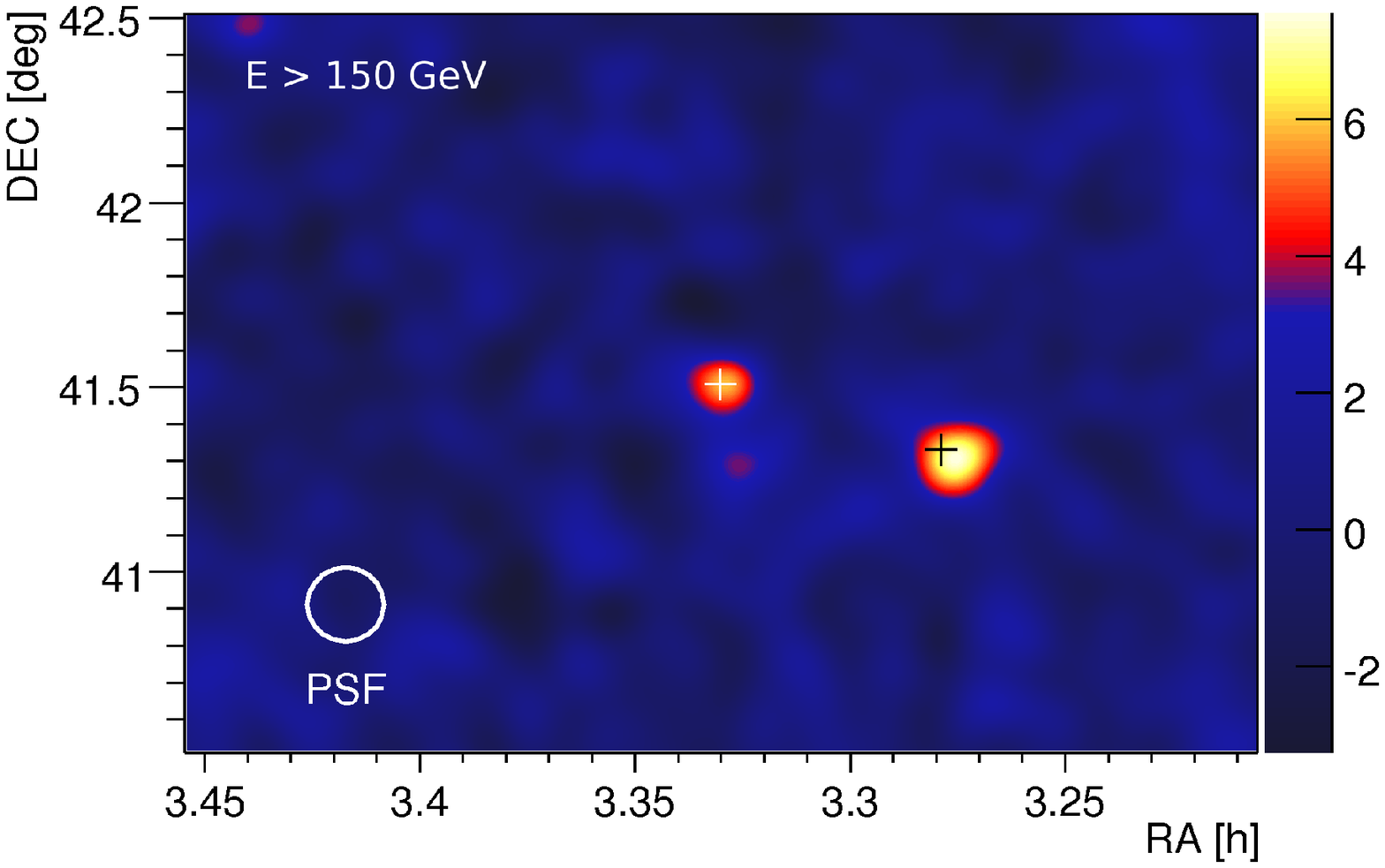}
\includegraphics[width=0.48\textwidth]{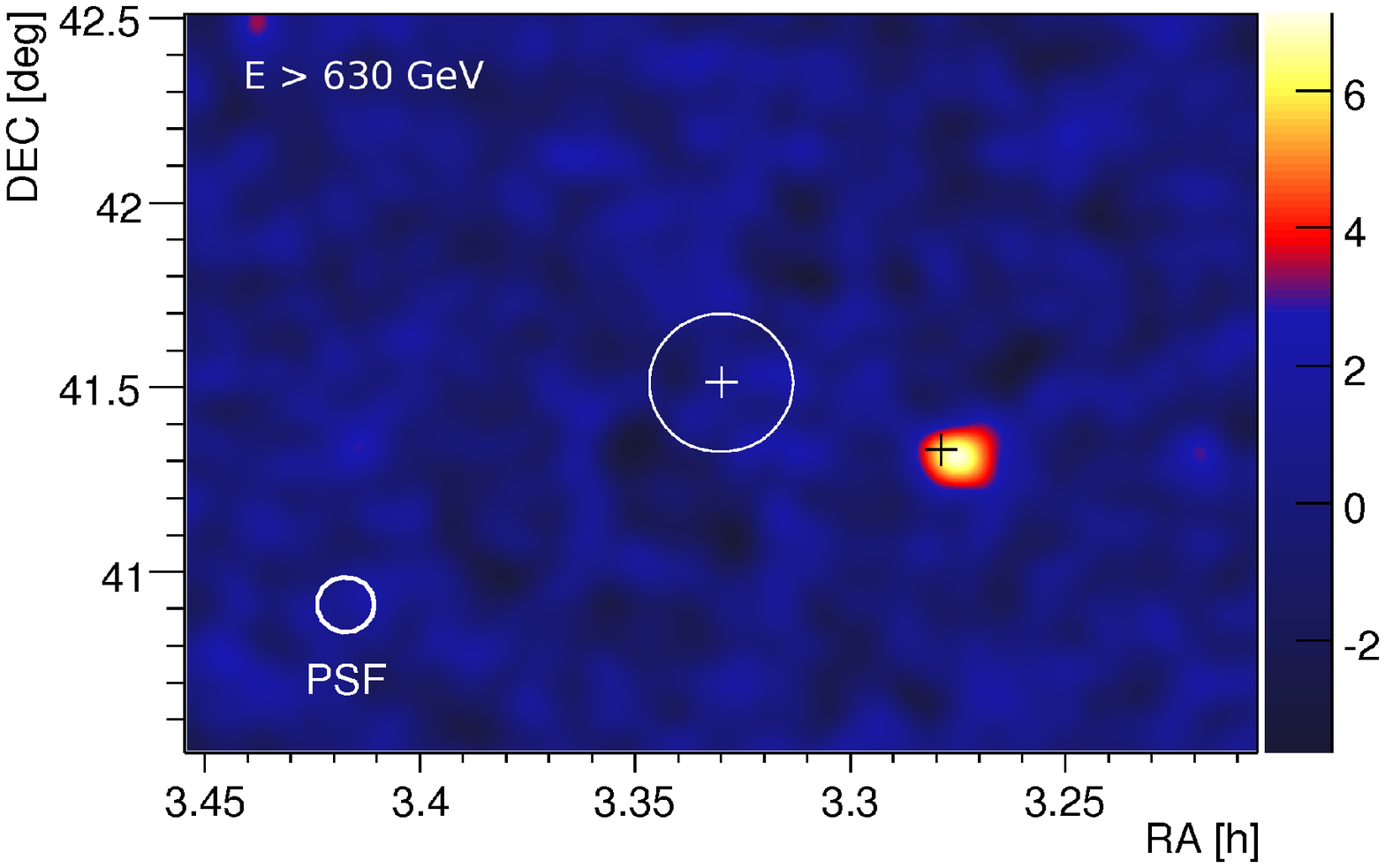}
\caption{
Left panel: significance skymap of the Perseus cluster above 150~GeV. The positions of NGC~1275 (white cross), which coincides with the cluster center, and the head-tail radio galaxy IC~310 (black cross) are marked. Right panel: significance skymap of the Perseus cluster region above 630~GeV. The central white circle marks the 0.15$^\circ$ region used to derive constraints (see main text for details). 
The quoted energy thresholds are obtained assuming a spectral index of 4 as observed in NGC~1275.
}
\label{skymap}
\end{figure*}

One of the most important questions in studies of nonthermal cluster emission is
the origin of giant radio halos in merging clusters and radio mini-halos in
cool-core clusters and whether they have a common physical origin. Hence it is
instructive to compare the prospects for detecting VHE gamma-ray emission from
two representative clusters of each model class, namely Coma and Perseus.
First, assuming universality of the gamma-ray spectrum, Coma is expected to be
fainter than Perseus by approximately a factor of 3.4. This is mainly because of
the lower central gas density in Coma \citep{2011PhRvD..84l3509P}. Second, the
Coma X-ray emission is more extended than in Perseus by the ratio of the
half-flux radii of $0.18^\circ/0.11^\circ =1.6$
\citep{2010MNRAS.409..449P}. Since the sensitivity of Imaging Atmospheric
Cherenkov Telescopes (IACTs) drops almost linearly with source extension for the
angular scales considered here, this makes it even more challenging to detect
the Coma cluster. Third, the CR spectral index is expected to steepen due to CR
transport processes such as streaming and momentum-dependent diffusion
\citep{2011A&A...527A..99E}. The authors of that paper indicate that these
mechanisms alleviate some tension between the hadronic model of radio halos
  and their observations, such as Coma. The following line of arguments shows
that this third point slightly favors Coma as a target source, but the sum of
all points clearly favors Perseus:
\begin{itemize}
\item Coma should have a central magnetic field strength of $B_{0} =
  4.7^{+0.7}_{-0.8}\,\mu$G according to Faraday RM studies
  \citep[e.g.,][]{2010A&A...513A..30B}. While GHz-radio observations probe
  2.5~GeV electrons that are produced in hadronic interactions of 40~GeV CR
  protons, gamma-rays at 200 GeV are from 1.6 TeV CRs: hence, for Coma, we have
  to extrapolate a factor of 50 in energy to connect the CRs probed by radio
  halo observations with those probed by IACTs and we expect a weak bias due to
  possible CR spectral steepening.
\item For Perseus, we expect a stronger central magnetic field ($\sim 20\,
  \mu$G, see discussion in Sect.~\ref{sec:B}) from adiabatic magnetic field
  compression during the formation of the cool core. While GHz-radio
  observations probe 1.2~GeV electrons and the 20~GeV CR protons that generate
  those electrons hadronically, the MAGIC gamma-ray constraints at 1~TeV
  correspond to 8~TeV CR protons (since the detection of VHE gamma-ray emission
  from the central AGN, NGC~1275, renders it difficult to observe diffuse
  emission at lower energies). Hence the CRs probed by the radio observations
  and those probed by MAGIC are a factor of 400 different in energy. A possible
  CR spectral steepening due to, e.g., momentum-dependent CR diffusion would
  induce a larger bias in the CR pressure and magnetic field
  constraints. Assuming a change in the spectral index by 0.2 between 20 GeV and
  8~TeV implies a decrease of the VHE gamma-ray flux by a factor of 3.3 --- 1.6
  times larger than the corresponding decrease of flux for Coma (over the eight
  times smaller energy range).
\end{itemize}

In summary, it appears that Perseus is the most promising target for
detecting CR-induced gamma-ray emission and hence we chose this cluster for our
deep VHE gamma-ray campaign. Assuming that the energy dependence of the CR
transport and the associated spectral steepening is representative for clusters,
we would need ten times longer integrations to detect gamma-ray emission from
Coma in comparison to Perseus.  However we stress that cool-core and non-cool
core clusters are complementary and equally deep or deeper observations of
merging clusters that host giant radio halos --- such as Coma --- are needed. We
also emphasize the need to theoretically model how the variation of the CR
spectral index due to energy dependent CR transport processes, e.g., CR
diffusion, depends on the mass and dynamical state of the cluster. This will be
critical for modeling the gamma-ray brightness of clusters and how to
interpret the underlying CR physics.

\section{MAGIC observations, analysis, and results}
\label{sec:obs}

The MAGIC telescopes are two 17\,m dish IACTs located at the Roque de los
Muchachos observatory ($28.8^\circ$N, $17.8^\circ$W, 2200~m a.s.l.), on the
Canary Island of La Palma. Since the end of 2009 the telescopes are working
together in stereoscopic mode which ensures a sensitivity of $\sim 0.7\%$ of the Crab
Nebula flux above approximately $600$~GeV in 50 hr of observations
\citep{2011arXiv1108.1477M}. For the energies of interest here (i.e., above 600
GeV), the point spread function (PSF), defined as a 2-dimensional Gaussian, has
a $\sigma\simeq0.06^\circ$.

The Perseus cluster region was observed by the MAGIC telescopes from
October 2009 to February 2011 for a total of about $99$~hr. During the
October 2009 - February 2010 campaign ($45.3$~hr), the data were taken in
the so called soft-stereo trigger mode with the first telescope (MAGIC-I) trigger working in
single mode and the second telescope (MAGIC-II) recording only events triggered by
both telescopes. The soft-stereo trigger mode may result in slightly degraded 
performance at the lowest energies with respect to \citet{2011arXiv1108.1477M}, 
but has a negligible impact at the energies of interest here.
During the August 2010 - February 2011 campaign ($53.6$~hr), data
were taken in the standard full-stereo trigger mode (where events are triggered simultaneously by both telescopes). 
Observations were performed in wobble mode \citep{1994APh.....2..137F} tracking positions $0.4^\circ$ from
the cluster center, at zenith angles ranging from 12$^\circ$ to 36$^\circ$.

The selected data sample consists of $84.5$~hr of effective observation time. The
data quality check resulted in the rejection of about $14.4$~hr of data,  
mainly due to non-optimal atmospheric conditions. The standard MAGIC
stereo analysis chain was used for calibration and image cleaning (see
\citealp{2011arXiv1108.1477M} for details). In order to combine data taken with
different trigger modes, we applied a high cut on the shower image size (total signal
inside the image). Only events with image size above 150~photo-electrons in
both telescopes were kept (the standard analysis uses 50~photo-electrons cuts).
With this cut, the rate and image parameter distributions of the background
events are compatible between the two samples.

The left panel of Fig.~\ref{skymap} shows the significance skymap obtained
with this analysis with an energy threshold of 150\,GeV.  The two AGNs detected
during the campaign can clearly be seen (NGC~1275 in the center and IC~310 in
the lower-right part).

In this work we limit our attention to pion-decay gamma-ray emission resulting
from hadronic CR interactions with thermal protons of the ICM.  Cosmological
simulations of \citet{2010MNRAS.409..449P} suggest that the spectral energy
distribution of gamma-rays follows a power-law, $F\propto E^{-\alpha}$, with a
spectral index of $\alpha=2.2$ at the energies of interest here
\citep{2010ApJ...710..634A}. The simulated signal is extended
with approximately 60$\%$ of the emission coming from a region centered on
NGC~1275 with a radius of $0.15^\circ$ (see Fig.~13 of
\citealp{2010MNRAS.409..449P}). The emission from NGC~1275 is dominant below
about 600~GeV and with a spectral index of about 4 \citep{2011arXiv1112.3917M}. 
Therefore, since the expected CR-induced signal is much harder than the
measured NGC~1275 spectrum, we expect it to appear at higher energies with no
break or cut-off in the energy range covered by MAGIC. Hence, we limit the data
analysis to energies $>630$~GeV for which the NGC~1275 signal vanishes \citep{2011arXiv1112.3917M}. 
The right panel of Figure~\ref{skymap} shows the significance
skymap above 630~GeV. In contrast to NGC~1275, the spectrum of IC~310 is very
hard and remains detectable above 600 GeV \citep{2010ApJ...723L.207A}. IC~310 is
$\sim 0.6^\circ(\simeq 10$~PSF) away from the cluster center, far enough so that
its emission does not leak into the signal region. However, if not explicitly
accounted for, it could affect the estimated background. Here, the background is
measured with three off-source positions at 0.4$^\circ$ from the the camera
center and $>0.28^\circ$ away from IC~310. This distance guarantees that there
is no contamination from IC~310.

\begin{figure}
\centering
\includegraphics[width=0.49\textwidth]{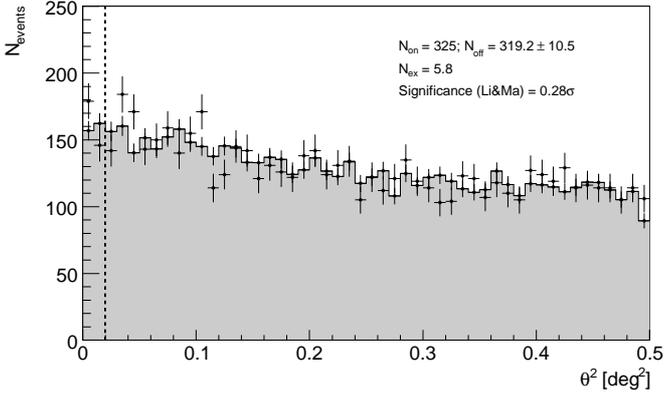}
\caption{
$\theta^2$ distribution above 630~GeV of the signal (data points) and background (gray filled region) at the cluster center. 
The dashed line represents the signal region cut within which the excess event number is estimated. Here it corresponds to $0.02$~deg$^2$.
The quoted energy threshold is obtained assuming a spectral index of 4 as observed in NGC~1275.
}
\label{thetaplot}
\end{figure}

In Fig.~\ref{thetaplot}, we show the $\theta^2$ distribution\footnote{The
  $\theta^2$ is the squared angular distance between the arrival direction of
  events and the nominal source positions (see
  e.g.~\citealp{1997APh.....8....1D}).} of the signal and background at the
cluster center for energies above $630$~GeV.  Since the emission is expected to
be more extended than the MAGIC PSF we must optimize the cut in the $\theta^2$
distribution which defines the signal region. We take the following approach.
First, we estimate the $\theta^2$ distribution of the expected signal
considering the simulated surface brightness of the Perseus cluster emission
smoothed with the MAGIC PSF.  Second, we estimate the background level from our
data.  Finally, we calculate the significance for signal detection (according to
the Eqn.~17 in \citealp{1983ApJ...272..317L}) as a function of the $\theta^2$
cut.  The derived optimal $\theta^2$ cut was close to $0.02$\,deg$^2$. As shown
in Fig.~\ref{thetaplot}, we find no significant excess within 0.02\,deg$^2$.

Since we did not detect a signal, we derived integral flux upper limits (ULs) for
several energy thresholds. The effective area of MAGIC is calculated using
point-like source Monte Carlo simulation and the ULs have to be corrected to take
into account the expected source extension. To calculate this correction factor,
we compare the fraction of the total events inside the signal region for a
point-like source and for the expected CR induced signal.  Therefore, the
presented ULs can be compared with the theoretical expectations for the region
within a radius of $0.15^\circ$. The UL estimation is performed using the Rolke
method \citep{2005NIMPA.551..493R} as in \cite{2011JCAP...06..035A} with a
confidence level of $95$\% and a total systematic uncertainty of
$30$\%.
In Table~\ref{tab:int_UL}, we present the integral
flux ULs above specified energy thresholds calculated for a spectral index of
$2.2$. We also give the point-like ULs, significance and ULs in number of events.
The integral ULs for energies above E$_{\rmn{th}}=1$~TeV corresponds to the best sensitivity 
for sources with spectral index $2.2$ and it is the most 
constraining value; for this reason we will adopt this UL for the discussion. For purposes that 
will be clear in the following, we also recalculate the $0.15^\circ$ integral flux UL
above $1$~TeV for spectral indexes of $2.1$, $2.3$ and $2.5$ which are
$1.37$, $1.38$ and $1.39\times10^{-13}$~cm$^{-2}$~s$^{-1}$, respectively.

\begin{table}[t]
\begin{center}
  \caption{\label{tab:int_UL} Integral flux ULs.
    }
\begin{tabular}{ccccc}
\hline\hline
\phantom{\Big|}
E$_{\mathrm{th}} [\mathrm{GeV}]$ & $\sigma_{\rmn{LiMa}}^{\rmn{PL}}$ & N$_{\rmn{UL}}^{\rmn{PL}}$ & F$_{\rmn{UL}}^{\rmn{PL}}$ & F$_{\rmn{UL}}^{0.15^{\circ}}$\\
\hline\\[-0.5em]
630    &   0.59$^a$  &  84.7  &   2.93  &  3.22\\
1000  &   0.15  &  41.4  &  1.25  & 1.38\\
1600  &   0.33  &  38.7  &  1.07  &  1.18\\ 
2500  &   0.38  &  28.8  &  0.79  &  0.87 \\  
\hline
\end{tabular}
\end{center}
\small{{\bf Note.} Integral flux ULs F$_{\rmn{UL}}$ for a power-law gamma-ray spectrum with spectral index $2.2$ above a given energy threshold
E$_{\mathrm{th}}$, both for a point-like source (PL) and for a $0.15^\circ$ extended region, in units of $10^{-13}$~cm$^{-2}$~s$^{-1}$. We additionally show the corresponding significance $\sigma_{\rmn{LiMa}}^{\rmn{PL}}$ and ULs in number of events N$_{\rmn{UL}}^{\rmn{PL}}$ (before applying the source extension correction).}
$^a$ \small{Note that the significance reported for E$_{\rm{th}}=630$~GeV is slightly different than in figure~\ref{thetaplot} because different cuts were applied. 
In particular, while computing the $\theta^2$ distribution we used a hard gamma-ray selection cut, which is normally adopted for detection purpose, in computing the flux ULs we used softer cuts in order to reduce the systematic errors.
} 
\end{table}

\section{Implications for cosmic rays and magnetic fields}
\label{sec:CR}

We follow different approaches in constraining the CR pressure distribution in
the Perseus cluster with the MAGIC ULs. This enables us to explore the
underlying plasma physics that produce the CR distribution and hence reflect the
Bayesian priors in the models \citep[see][for a
discussion]{2011PhRvD..84l3509P}. Our approaches include (1) a simplified
analytical approach that assumes a constant CR-to-thermal pressure and a
momentum power-law spectrum (Sect.~\ref{sec:simplified}), (2) an analytic model
of CRs derived from cosmological hydrodynamical simulations of the formation of
galaxy clusters (Sect.~\ref{sec:sim}), and (3) the use of the observed
luminosity and surface brightness profile of the radio mini-halo in Perseus to
place a lower limit on the expected gamma-ray flux in the hadronic model of
radio mini-halos --- where the radio-emitting electrons are secondaries from CR
interactions (Sect.~\ref{sec:min}).  This provides a minimum CR pressure which,
using tight gamma-ray limits/detections, checks the hadronic model of the
formation of radio mini-halos. (4) Alternatively, by constructing a CR
distribution that is just allowed by the flux ULs, and requiring the
model to match the observed radio mini-halo data, we can derive a lower limit on
the magnetic field strength (Sect.~\ref{sec:B}). Note, however, that this limit assumes that the
observed synchrotron emission is produced by secondary electrons resulting from
hadronic CR interactions.

\subsection{Simplified analytical CR model}
\label{sec:simplified}

Here, we adopt a simplified analytical model that assumes a power-law CR
momentum spectrum, $f \propto p^{-\alpha}$, and a constant CR-to-thermal pressure ratio,
i.e., we adopt the {\em isobaric model of CRs} following the approach of
\citet{2004A&A...413...17P}. To facilitate comparison with earlier analytical
work in the literature, we do not impose a low-momentum cutoff on the CR
distribution function, i.e., we adopt $q=0$ (this assumption can be easily
generalized to an arbitrary $q$ using e.g., Fig. 1 of
\citealt{2004A&A...413...17P}). Since {\it a priori} the CR spectral
index\footnote{The hadronic interaction physics guarantees that the CR spectral
  index coincides with that from pion-decay gamma-ray emission.}, $\alpha$, is
unconstrained, we vary it within a plausible range of $2.1<\alpha<2.5$. The
  central value of this spectral range is compatible with the radio spectral
index in the core of the Perseus radio mini-halo of
$\alpha_\nu=\alpha/2\sim1.25$ \citep{Sijbring1993}:\footnote{The reported radial
  spectral steepening of the radio mini-halo emission
  \citep{2002A&A...386..456G} could be an observational artifact owing to a poor
  signal-to-noise ratio in the outer core of the cluster and the ambiguity in
  determining the large scale Fourier components owing to nonuniform coverage of
  the Fourier plane and missing short-baseline information: the so-called
  ``missing zero spacing''-problem of interferometric radio observations.  By
  comparing the spectral index distribution of the three radio maps \citep[at 92
  cm, 49 cm, and 21 cm,][]{Sijbring1993}, radial spectral flattening depending
  on the chosen radial direction seems possible.}  assuming a central magnetic
field strength of 20~$\mu$G, appropriate for cool-core clusters, the CR protons
responsible for the GHz radio emitting electrons have an energy of $\sim20$ GeV
and are $\sim 400$ times less energetic than the CR protons with an energy of 8
TeV that are responsible for gamma-ray emission at 1~TeV.  This is consistent
with the concavity in the universal CR spectrum that connects the steep-spectrum
low-energy CR population around GeV energies ($\alpha\simeq 2.5$) with the
harder CR population at TeV energies ($\alpha\simeq 2.2$)
\citep{2010MNRAS.409..449P}.  To model the thermal pressure, we adopt the
measured electron temperature and density profiles for the Perseus cluster
\citep{2003ApJ...590..225C}. The temperature profile has a dip in the central
cool core region and otherwise a constant temperature of $kT = 7$ keV.  The
density profile is hybrid in the sense that it combines {\em Einstein} X-ray
observations on large scales with high-resolution {\em XMM-Newton} observations
of the cluster core \citep{2003ApJ...590..225C}.

Table \ref{tab:XCR} shows the resulting constraints on the CR-to-thermal pressure ratio,
$X_{\CR} = \expval{P_{\CR}}/\expval{P_\rmn{th}}$, averaged within the virial
radius, $R_\rmn{vir}=2$~Mpc, that we define as the radius of a sphere enclosing
a mean density that is 200 times the critical density of the Universe.  The
inferred constraints on $X_\CR$ strongly depend on $\alpha$ due to the
comparably large lever arm from GeV-CR energies (that dominate the CR pressure)
to CR energies at 8 TeV. Using the integral flux UL above 1~TeV, 
we  constrain $X_\CR$ to be between 0.77\% and 11.6\% (for $\alpha$
varying between 2.1 and 2.5).  For a spectral index of 2.2, favored by the
simulation-based model of \citet{2010MNRAS.409..449P} at energies $> 1$~TeV, we
obtain  $X_\CR<1.1\%$.

\begin{table}[t]
\begin{center}
  \caption{\label{tab:XCR} Constraints on CR-to-thermal pressure ratio.}
\begin{tabular}{ccccc}
\hline\hline
\phantom{\Big|}
$\alpha$ & $X_{\CR,\rmn{max}}$ [\%] & $X_{\CR,\rmn{min}}$ [\%] & 
$X_{\CR,\rmn{max}}/X_{\CR,\rmn{min}}$ & 
$F_{\gamma,\rmn{min}}^a$\\
\hline\\[-0.5em]
2.1 & 0.77 & 0.42 & ~1.8 & 7.4 \\
2.2 & 1.12 & 0.35 & ~3.2 & 4.3 \\
2.3 & 2.17 & 0.38 & ~5.7 & 2.4 \\
2.5 & 11.6 & 0.67 & 17.3 & 0.8 \\[0.5em]
\hline
\end{tabular}
\end{center}
{\bf Note.} Constraints on CR-to-thermal pressure ratio in the Perseus
    cluster core, $X_{\CR,\rmn{max}}$, which is assumed to be constant in the
    simplified analytical model. Those constraints are compared to minimum
    CR-to-thermal pressure ratios, $X_{\CR,\rmn{min}}$, and minimum gamma-ray
    fluxes in the hadronic model of the Perseus radio mini-halo.
$^a$ Minimum gamma-ray flux in the
hadronic model of the Perseus radio mini-halo, $F_{\gamma,\rmn{min}} (> 1\,\rmn{TeV})$, in units of $10^{-14}$~cm$^{-2}$~s$^{-1}$.
\end{table}

\subsection{Simulation-inspired CR model}
\label{sec:sim}
For a more realistic approach, we turn to cosmological hydrodynamical
simulations.  These have considerable predictive power, e.g., they calculate the
CR spectrum self-consistently rather than leaving it as a free parameter, and
permit tests of various assumptions about the underlying CR physics.  We adopt
the universal spectral and spatial CR model developed by
\citet{2010MNRAS.409..449P} that uses only a density profile as input which can
be inferred from cosmological simulations or X-ray observations. This analytic
approach models the CR distribution, and the associated radiative emission
processes from hadronic interactions with gas protons, from radio to the
gamma-ray band. The gamma-ray emission from decaying neutral pions dominates
over the IC emission from primary and secondary electrons above 100 MeV in
clusters \citep[e.g.,][]{2008MNRAS.385.1242P}.  This analytic formalism was
derived from high-resolution simulations of clusters that included radiative
hydrodynamics, star formation, supernova feedback, and followed CR physics
spectrally and spatially by tracing the most important injection and loss
processes self-consistently while accounting for the CR pressure in the equation
of motion \citep{2006MNRAS.367..113P,2007A&A...473...41E,2008A&A...481...33J}.
The results are consistent with earlier numerical results within their range of
applicability regarding the overall characteristics of the CR distribution and
the associated radiative emission processes
\citep{2000A&A...362..151D,2001ApJ...559...59M,2003MNRAS.342.1009M,
  2007MNRAS.378..385P,2008MNRAS.385.1211P,2008MNRAS.385.1242P}.

There are (at least) two major uncertainties in modeling the CR physics that
significantly affect the resulting spatial and spectral CR distribution (and may
impact the universality of the CR populations across different cluster masses):
the CR acceleration efficiency and the microscopic CR transport relative to the
thermal plasma.  We will discuss each process separately. The overall
normalization of the CR and gamma-ray distribution scales with the maximum
acceleration efficiency at structure formation shocks. Following recent
observations \citep{2009Sci...325..719H} and theoretical studies
\citep{2005ApJ...620...44K} of supernova remnants, we adopt an optimistic but
realistic value of this parameter and assume that 50\% of the dissipated energy
at strong shocks is injected into CRs while this efficiency rapidly decreases
for weaker shocks. The vast majority of internal formation shocks (merger and
flow shocks) are weak with Mach numbers $M\lesssim3$
\citep[e.g.,][]{2003ApJ...593..599R, 2006MNRAS.367..113P} which we presume have
low CR acceleration efficiencies and do not contribute significantly to the CR
population in clusters. Instead, strong shocks during the cluster formation
epoch and strong accretion shocks at the present time (at the boundary of voids
and filaments/super-cluster regions) dominate the injection of CRs which are
adiabatically transported throughout the cluster.  These models provide a
plausible UL for the CR contribution from structure formation shocks in galaxy
clusters which can be scaled with the effective acceleration efficiency. Other
possible CR sources, such as starburst-driven galactic winds and AGNs, could
increase the expected gamma-ray yield, but have been
neglected.\footnote{Adiabatic CR losses during the expansion phase of the
  galactic winds imply a negligible contribution to the CR energy density of the
  ICM today \citep{2010ApJ...710..634A} but integrated over cluster history,
  they may contribute an interesting seed population that however needs to be
  re-accelerated. It is currently unclear whether AGN jets are powered by the
  Poynting flux or by the kinetic energy of hadrons
  \citep[e.g.,][]{1998MNRAS.293..288C, 1998tx19.confE.407H,
    2000ApJ...534..109S}. Even if there was a non-negligible CR ion component,
  it would probably be confined to dilute radio lobes that rise almost
  unperturbed in the cluster atmosphere to the less dense outer cluster regions
  \citep{2007ARA&A..45..117M} where these lobes may be destroyed by shocks and
  dissolved within the ICM \citep{2002MNRAS.331.1011E,
    2011ApJ...730...22P}. Since the CRs are then released into an ambient ICM
  with a small density, this implies a negligible contribution to the cluster
  gamma-ray emission due to pion decay.}

These cosmological simulations only consider advective transport of CRs by
turbulent gas motions which produces a centrally enhanced profile.  However,
active CR transport such as CR diffusion and streaming flattens the CR radial
profile, producing a spatially constant CR number density in the limiting case.
Although the CR streaming velocity is larger than typical advection velocities
in a galaxy cluster, it becomes comparable to or lower than this only for
periods with trans- and supersonic cluster turbulence during a cluster
merger. As a consequence a bimodality of the CR spatial distribution results
with merging (relaxed) clusters showing a centrally concentrated (flat) CR
energy density profile \citep{2011A&A...527A..99E}.  This produces a
bimodality of the diffuse radio and gamma-ray emission of clusters, since more
centrally concentrated CRs will find higher target densities for hadronic CR
proton interactions \citep{2011A&A...527A..99E}. Thus, relaxed clusters could
have a reduced gamma-ray luminosity by up to a factor of five. The goal of this
and future searches for gamma-ray emission from clusters is to constrain the
acceleration physics and transport properties of CRs.

In the following, we use two different CR models. In our optimistic CR model
({\em simulation-based analytics with galaxies}), we calculated the cluster
total gamma-ray flux within a given solid angle. In contrast, we cut the
emission from individual galaxies and compact galactic-sized objects in our more
conservative model ({\em simulation-based analytics without galaxies}). In
short, the ICM is a multiphase medium consisting of a hot phase which attained
its entropy through structure formation shock waves dissipating gravitational
energy associated with hierarchical clustering into thermal energy. The dense,
cold phase consists of the true interstellar medium within galaxies (which is
modelled with an effective equation of state following
\citealt{2003MNRAS.339..289S}) and at the cluster center as well as the
ram-pressure stripped interstellar medium.  These cold dense gas clumps
dissociate incompletely in the ICM due to insufficient numerical resolution as
well as so far incompletely understood physical properties of the cluster
plasma.  All of these phases contribute to the gamma-ray emission from a
cluster. To assess the bias associated with this issue, we performed our
analysis with both limiting cases bracketing the realistic case.

\begin{figure}
\centering
\includegraphics[width=0.5\textwidth]{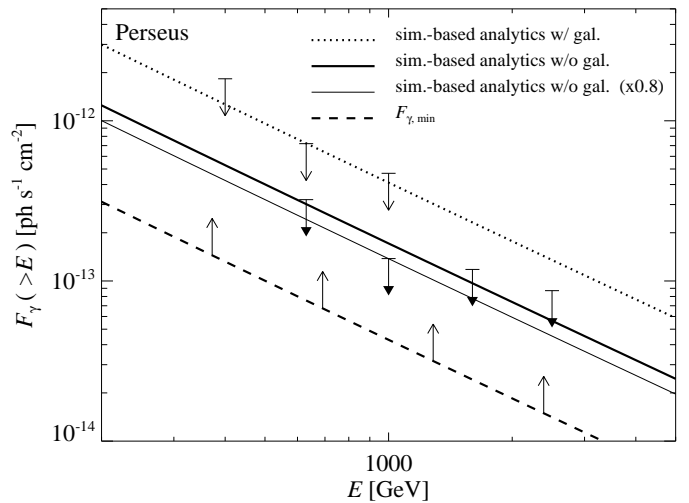}
\caption{Integral flux ULs of the single telescope observations (point-like ULs,
  upper arrows; \citealp{2010ApJ...710..634A}) and of this work (solid arrows;
  Table~1) are shown. We compare ULs with the simulated integrated spectra of
  the gamma-ray emission from decaying neutral pions that result from hadronic
  CR interactions with the ICM in the Perseus cluster coming from within a
  radius of $0.15^\circ$ around the center.  The conservative model without
  galaxies (solid line) is shown together with the model with galaxies (dotted
  line). The flux UL at 1~TeV is a factor of 1.25 below the conservative model
  and hence constrains a combination of CR shock acceleration efficiency and CR
  transport processes. We also show the minimum gamma-ray flux estimates for the
  hadronic model of the Perseus radio mini-halo (dashed line with minimum flux
  arrows) using the universal gamma-ray spectrum resulting from pion decay
  \citep{2010MNRAS.409..449P} and adopting the spectral index, 2.2, of
  this spectrum around TeV energies.}
\label{fig:spectrum}
\end{figure}

We again adopt the hybrid electron density and temperature profile from X-ray
observations of the Perseus cluster \citep{2003ApJ...590..225C}\footnote{While
  the X-ray data show a complicated ICM structure including cavities blown by
  the central AGN, the use of deprojected, spherically averaged gas density and
  temperature profiles can be justified as follows.  The inhomogeneities of the
  ICM can be quantified statistically using a clumping factor $C_\rho =
  \sqrt{\expval{\rho^2}} / \expval{\rho}$ that estimates the relative bias
  factor to true density profile. However, since we are interested in the
  CR-induced gamma-ray emissivity which also scales with the square of the gas
  density (modulo a dimensionless CR-to-gas ratio that we quantify by using CR
  simulations) our modeling should inherit exactly the same bias as in
  X-ray-inferred density profiles. We emphasize that the overall X-ray
    emission is smooth with fluctuations on the level $<20-25\%$
    \citep{2011MNRAS.418.2154F}. Those substructures are only amplified by
    unsharp mask imaging.  Moreover, the X-ray cavities (interpreted as AGN
    bubbles) do not correlate with structures in the radio mini-halo and only
    make up a small fraction of the radio-emitting volume ($< 10^{-3}$, since
    they only subtend a linear scale of $< 10\%$ relative to the radio emission,
    \citealt{2011MNRAS.418.2154F}), justifying our procedure.}, modifying the
temperature profile in the outer cluster regions beyond $0.2 R_{200} \simeq
400\,\rmn{kpc}$.  This produces the characteristic decline toward the cluster
periphery in accord with a fit to the universal profile derived from
cosmological cluster simulations \citep{2010MNRAS.409..449P,
  2007MNRAS.378..385P} that also follows the behavior of a nearby sample of deep
{\em Chandra} cluster data \citep{2005ApJ...628..655V}. While this modification
has little influence on the expected gamma-ray emission (in projection onto the
core region) as the densities drop considerably in these regions, the resulting
profiles for the CR-to-thermal pressure ratio are changed as we will discuss below. We
estimate the spatial and spectral distribution of CRs using the model by
\citet{2010MNRAS.409..449P}. Figure \ref{fig:spectrum} shows the expected
spectrum for Perseus within an aperture of radius 0.15$^\circ$. The MAGIC limit
for $E_\gamma>1$~TeV falls below the flux level of the conservative model (that
does not take into account the emission from galaxies\footnote{Note that the
  models ``with galaxies'' and ``without galaxies'' differ from those used in
  the first MAGIC paper on Perseus observations with a single telescope
  \citep{2010ApJ...710..634A}. Previously, the models were based on the
  simulated cool core cluster g51 (which is similar in morphology to Perseus)
  while here we use the universal CR model --- which is based on high-resolution
  simulations of 14 galaxy clusters that span almost two decades in mass --- to
  compute the expected pion decay emission \citep{2010MNRAS.409..449P}. The
  spectral normalization (between the two models) therefore increased from 1.5
  to 2.4 and the gamma-ray flux in the model ``without galaxies'' used here
  differs by 10\%. }) by 20\%; thereby constraining assumptions about the
adopted CR physics in the simulations and the resulting CR pressure.

\begin{figure}
\centering
\includegraphics[width=0.5\textwidth]{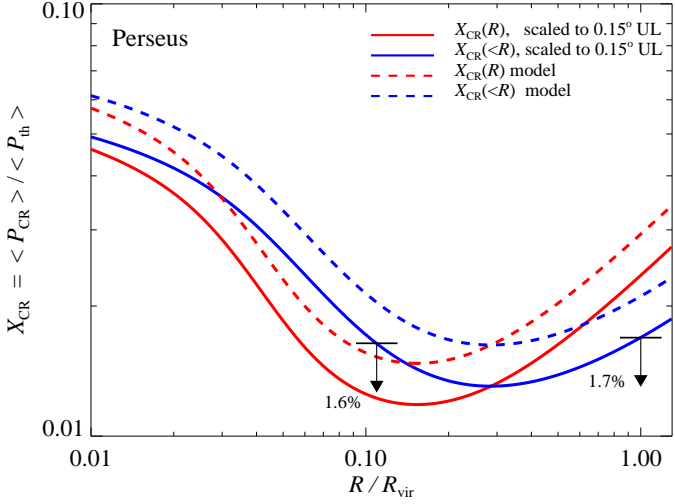}
\caption{CR-to-thermal pressure ratio, $X_{\CR} = \expval{P_{\CR}}/\expval{P_\rmn{th}}$
  at the radial distance from the Perseus cluster center, $R/R_\rmn{vir}$ (red
  lines), and integrated up to $R/R_\rmn{vir}$ (blue lines), using the
  simulation-based analytical model of CRs \citep{2010MNRAS.409..449P}. The
  simulation model (dashed) is contrasted to a model that has
  been scaled by the MAGIC constraints obtained in this work (solid).}
\label{fig:XCR}
\end{figure}

Figure~\ref{fig:XCR} shows the CR-to-thermal pressure ratio, $X_{\CR} =
\expval{P_{\CR}}/\expval{P_\rmn{th}}$ as a function of radial distance from the
Perseus cluster center, in units of the virial radius, $R_\rmn{vir}$.  To
compute $P_\CR$, we assume a low-momentum cutoff of the CR distribution of $q =
0.8\,m_p c$ found in cosmological simulations by
\citet{2010MNRAS.409..449P}. This cutoff derives from the high cooling rates for
low energy CRs due to Coulomb interactions with the thermal plasma.  The ratio
$X_{\CR}$ {\em rises toward the outer regions} because of the higher CR
acceleration efficiency in the peripheral strong accretion shocks compared to
the weak central flow shocks. Adiabatic compression of a mixture of CRs and
thermal gas reduces the CR pressure relative to the thermal pressure due to the
softer CR equation of state. {\em The strong increase of $X_{\CR}$ toward the
  core} is a remnant of the formation of the cool core in Perseus. During this
transition, the mixed gas of CRs and thermal particles has been adiabatically
compressed. While the thermal gas radiates on a comparatively fast thermal
bremsstrahlung timescale, the long hadronic interaction time scale for energetic
CRs ensures an accumulation of this population, thus diminishing the thermal
pressure support relative to that in CRs.

The MAGIC flux limits constrain $X_{\CR}$ of the simulation-based analytical
model to be less than 1.6\% within 0.15$^\circ$ ( 200 kpc). Assuming this
spatial CR profile yields a CR-to-thermal pressure ratio $< 1.7\%$ within
$R_\rmn{vir}\simeq 2$~Mpc and $< 5\%$ within 20~kpc. When scaling the $X_{\CR}$
profile to the flux ULs, we used the conservative model of
\citet{2010MNRAS.409..449P} that neglects the contribution of cluster galaxies
since those increase only the gamma-ray yield but do not contribute to the CR
pressure within the ICM. The $X_{\CR}$-limit within the virial radius is larger
by a factor of 1.5 than a pure power-law spectrum (see
Sect.~\ref{sec:simplified}) because the concave curvature of the simulated
spectrum accumulates additional pressure toward GeV energies relative to a pure
power-law.

With these gamma-ray flux ULs, we constrain the CR physics in galaxy clusters:
this either limits the maximum acceleration efficiency of CRs at strong
structure formation shocks to $<50\%$ or indicates possible CR streaming and
diffusion out of the cluster core region. The latter populates the peripheral
cluster regions and $X_\CR$ increases toward the cluster periphery at the
expense of a decrease of the central $X_\CR$ \citep{2011A&A...527A..99E}
compared to the simulation model. The X-ray morphology of the central region in
Perseus shows spiral structure in the density and temperature maps with an
anti-correlation of both quantities \citep{2011MNRAS.418.2154F}.  This resembles
sloshing motions after a past merger event, suggesting that Perseus is currently
relaxing. This interpretation is supported by recent magnetohydrodynamical
simulations \citep{2011arXiv1108.4427Z} which argue that magnetic fields are
draped at the contact discontinuity of the remnant cool core
\citep{2006MNRAS.373...73L, 2008ApJ...677..993D, 2010NatPh...6..520P} that is
sloshing in the potential of the parent halo. The abrupt drop in the radio
surface brightness map outside the spiral pattern \citep[see Fig. 8
in][]{2011MNRAS.418.2154F} may then be produced by a strong decrease of magnetic
field strengths outside the sloshing core. If CR streaming and diffusion out
  of the central core region is indeed correlated with a dynamical relaxation of
  a cluster after a merger event as suggested by \citet{2011A&A...527A..99E},
  this would render CR transport more plausible for explaining the smaller
gamma-ray flux relative to the simulation model. By the same token, it may argue
for a more extended gamma-ray emission signature than that seen in the radio,
further justifying the larger source extension, 0.15$^\circ$, that we adopted in
this work (which is twice that of the largest radial extend of the mini-halo of
0.075$^\circ$).

\subsection{Minimum gamma-ray flux}
\label{sec:min}

For clusters with radio (mini-)halos we can derive a minimum gamma-ray flux in
the hadronic model of radio (mini-)halos. A stationary distribution of CR
electrons loses all its energy to synchrotron radiation for strong magnetic
fields ($B \gg B_\rmn{CMB} \simeq 3.2 \mu$G, where $B_\rmn{CMB}$ is the
equivalent magnetic field strength of the cosmic microwave background (CMB) so
that $B_\rmn{CMB}^2/8\pi$ equals the CMB energy density). Hence the ratio of
gamma-ray to synchrotron flux becomes independent of the spatial distribution of
CRs and thermal gas \citep{Voelk:1989,Pohl:1994,2008MNRAS.385.1242P},
particularly with the observed synchrotron spectral index $\alpha_{\nu}=\alpha/2
\simeq 1$. This can be easily seen by considering the pion-decay induced
gamma-ray luminosity, $L_\gamma$, and the synchrotron luminosity, $L_\nu$, of a
steady state distribution of CR electrons that has been generated by hadronic CR
interactions,
\begin{eqnarray}
L_\gamma &=& A_\gamma \dps\int d V\, n_\CR n_\rmn{gas},\\
\label{eq:Lnu}
   L_\nu &=& A_\nu \dps\int d V\, n_\CR n_\rmn{gas} \frac{\dps
    \eps_B^{(\alpha_\nu+1)/2}}{\dps\eps_\rmn{CMB}+\eps_B}\\
  &\simeq&A_\nu \dps\int d V\, n_\CR n_\rmn{gas}\quad\rmn{for}~\eps_B \gg \eps_\rmn{CMB}.
\end{eqnarray}
Here, $\eps_B$ and $\eps_\rmn{CMB}$ denote the energy density of the magnetic
field and the CMB, respectively, $A_\gamma$ and $A_\nu$ are dimensional constants that
depend on the hadronic physics of the interaction \citep{2008MNRAS.385.1211P,
  2008MNRAS.385.1242P}.  Hence we can derive a minimum gamma-ray flux in the
hadronic model:
\begin{equation}
  \label{eq:Fmin}
  F_{\gamma,\rmn{min}} = 
  \frac{\dps A_\gamma}{\dps A_\nu}\frac{\dps L_\nu}{\dps 4\pi D_\rmn{lum}^2},
\end{equation}
where $L_\nu$ is the observed luminosity of the radio mini-halo and
$D_\rmn{lum}=77$~Mpc denotes the luminosity distance to Perseus.  Lowering the
magnetic field strength would require increasing the CR electron energy density
to reproduce the observed synchrotron luminosity and thus increases the
associated gamma-ray flux. The maximum emission radius of the radio mini-halo of
100 kpc corresponds to an angular size of 0.075$^\circ$, well within the MAGIC
PSF; hence $L_\nu$ does not need to be cut to match the angular region tested by
the MAGIC ULs.

Hence in this section, we assume $B \gg B_\rmn{CMB}$ everywhere in the
radio-emitting region, and (in order to obtain this minimum gamma-ray flux)
require a considerable drop in the CR distribution outside the radio
(mini-)halo. While we do not make any assumptions on how to realize such a
scenario, we discuss possible physical processes that could be causing it.
The large magnetic field can be realized by adiabatically compressing the
  magnetic field during the formation of the cool core.  The edge of the radio
emitting region can be caused by a region of predominantly toroidal magnetic
field which efficiently confines CRs to the central region. Such a magnetic
field configuration could be arranged by magnetic draping at the sloshing core
\citep{2006MNRAS.373...73L, 2008ApJ...677..993D, 2010NatPh...6..520P} or the
saturated (non-linear) state of the heat-flux driven buoyancy instability
\citep{2008ApJ...673..758Q, 2008ApJ...677L...9P}. The drop in CR number
  density outside this region can be realized by CR streaming to the peripheral
  cluster regions with substantially lower target densities that would imply a
  negligible contribution to the gamma-ray flux outside the radio mini-halo
  \citep{2011A&A...527A..99E}. The next generation of cluster simulations that
  couple CR physics self-consistently to magnetic fields and account for
  anisotropic transport processes of the thermal gas as well as CRs will be
  needed to scrutinize such a presented scenario.

In contrast to our previous analysis in \citet{2010ApJ...710..634A}, we derive
here the {\it absolute} minimum gamma-ray flux rather than the ``physical''
minimum gamma-ray flux that depends on the assumed magnetic field distribution
and requires an uncertain extrapolation into the outer regions of the
cluster. The density and pressure profile of Perseus as derived from X-ray
observations can both be described by double-beta profiles with an inner core
radius of 57 and 47~kpc, respectively \citep{2003ApJ...590..225C,
  2005A&A...430..799P}. Hence, there is only a small interval between $\sim50$
and $\sim100$ kpc, the maximum radius of the radio mini-halo, that is sensitive
to the outer CR and magnetic field profile. Only the extrapolation of that
behavior well beyond the radio mini-halo to the virial radius will determine
whether the energy density in the magnetic field becomes implausibly large in
comparison to that of the thermal gas, thereby possibly violating the energy
conditions.

The results for the minimum gamma-ray flux $F_{\gamma,\rmn{min}}(>1~\rmn{TeV})$
and the minimum CR-to-thermal pressure ratio, $X_{\CR,\,\rmn{min}} = X_\CR
F_{\gamma,\rmn{min}}/F_{\gamma,\rmn{iso}}$ are presented in Table \ref{tab:XCR}
and Fig. \ref{fig:spectrum} (assuming $\alpha=2.2$). Here,
$F_{\gamma,\rmn{iso}}$ is the gamma-ray flux in the {\em isobaric model of CRs}
introduced in Sect.~\ref{sec:simplified}.  The ratio of the maximum to minimum
CR pressures, $X_{\CR,\rmn{max}}/X_{\CR,\rmn{min}}$, varies between 1.8 and 17.3
for a spectral index between $2.1 \leq \alpha \leq 2.5$. For the spectral index
$\alpha=2.2$ of the universal gamma-ray spectrum around TeV energies,
the ratio is $X_{\CR,\rmn{max}}/X_{\CR,\rmn{min}}=3.2$. This puts the
long-sought gamma-ray detection of clusters, in particular for Perseus, within
the reach of deeper campaigns with the possibility of scrutinizing the hadronic
emission model of radio (mini-)halos.

\subsection{Implications for the cluster magnetic field}
\label{sec:B}

As shown in Sect.~\ref{sec:min}, an absolute lower limit on the hadronic model
gamma-ray emission comes from assuming high magnetic field strengths, $B\gg
B_\rmn{CMB}$, everywhere in the radio-emitting region. Using our UL on the
gamma-ray emission (and on $P_{\CR}$) we can turn the argument in
Sect.~\ref{sec:min} around to derive a lower limit on the magnetic field
strength needed to explain the observed diffuse radio emission within the
hadronic model \citep{2004MNRAS.352...76P}. Lowering the gamma-ray limit will
tighten (increase) the magnetic field limit. If this conflicts with magnetic
field measurements by means of other methods, e.g., Faraday RM, this would
challenge the hadronic model of radio (mini-)halos. The method that we use to
constrain the magnetic field, $B$, depends on the assumed spatial structure of
$B$ that we parametrize by
\begin{equation}
\label{eq:B}
B(r) = B_{0} \,\left(\frac{n_{\rmn{e}}(r)}{n_{\rmn{e}}(0)}\right)^{\alpha_B},
\end{equation}
as suggested by a cosmological MHD simulation of a cool core cluster
\citep{2008A&A...482L..13D}, which demonstrates a tight correlation of the
magnetic field with the ICM gas density and highlights the importance of cooling
processes in amplifying the magnetic field in the core of galaxy clusters up to
one order of magnitude above the typical amplification obtained for a purely
adiabatic evolution. Moreover, such a parametrization is favored by Faraday RM
studies \citep[][and references therein]{2010A&A...513A..30B,
  2011A&A...529A..13K}.

Generally, Faraday RM determinations of the magnetic field strength, for instance using   
background sources seen through clusters, depend intrinsically on the magnetic
correlation length. Recent Faraday RM studies yield estimates for the central
magnetic field of typically 5~$\mu$G for merging clusters \citep[][for
Coma]{2010A&A...513A..30B} and significantly higher values in cool core clusters
of around 16~$\mu$G \citep[][for Hydra A]{2011A&A...529A..13K}. For the Perseus
cluster, Faraday RMs are available only on very small scales
\citep{2006MNRAS.368.1500T}, i.e., a few tens of pc.  The RM estimates are
$\sim7000 \mbox{ rad m}^{-2}$ implying magnetic field strengths of $\sim25~\mu$G.
This assumes that the Faraday screen is situated within the ICM.  This location
is not, however, certain since variations of $10\%$ in the RM are observed on
pc-scales \citep{2002MNRAS.334..769T}, while ICM magnetic fields are expected to
be ordered on scales of a few kpc
\citep{2006MNRAS.368.1500T,2005A&A...434...67V}.

To proceed, we derived a deprojected radio emissivity profile.  We fit the
point-source subtracted, azimuthally averaged surface brightness profile at 1.38
GHz \citep{1990MNRAS.246..477P} with a $\beta$-model,
\begin{equation}
\label{beta}
 S_{\nu} (r_{\bot})= S_{0} \left[ 1 + \left( \frac{r_{\bot}}{r_{\rmn{c}}}\right)^{2}\right]^{-3\beta + 1/2},
\end{equation}
where $S_{0} = 2.3 \times 10^{-1}\,\rmn{Jy\,arcmin}^{-2}$, $r_{\rmn{c}} = 30$
kpc, and $\beta = 0.55$. This profile is valid within a maximum emission radius
of 100~kpc. An Abel integral  deprojection then provides the
radio emissivity distribution (see Appendix of \citealt{2004A&A...413...17P}),
\begin{equation}
\label{eq:Perseus:radio}
j_{\nu} (r) = \frac{S_{0}}{2\pi\, r_{\rmn{c}}}\,
\frac{6\beta - 1}{\left(1 + r^{2}/r_{\rmn{c}}^{2}\right)^{3 \beta}}\,
\mathcal{B}\left(\frac{1}{2}, 3\beta\right)
= j_{\nu,0} \left(1 + r^2/r_{\rmn{c}}^{2}\right)^{-3 \beta},
\end{equation}
where $\mathcal{B}$ denotes the beta function.  To constrain the
magnetic field, we proceed as follows:
\begin{enumerate}
\item Given a model for the magnetic field characterized by $\alpha_B$ and an
  initial guess for $B_0$, we determine the profile of 
  $X_{\CR}(r)$ such that the hadronically produced synchrotron emission 
  matches the observed radio mini-halo emission over the entire extent.
\item For this  $X_{\CR}(r)$ profile, we compute the pion-decay gamma-ray
  surface brightness profile, integrate the flux within a radius of
  $0.15^\circ$, and scale the CR profile o match the corresponding
  MAGIC UL. This scaling factor, $X_{\CR,0}$, depends on the CR
  spectral index, $\alpha$, (assuming a power-law CR population for simplicity),
  the radial decline of the magnetic field, $\alpha_{B}$, and the initial guess
  for $B_0$.
\item We then solve for $B_{0}$ requiring that it matches the observed
  synchrotron profile while fixing the profile of $X_{\CR}(r)$ determined in the
  previous two steps.  Note that for $B_{0} \gg B_{\rmn{CMB}}$ and a radio
  spectral index of $\alpha_{\nu}=1$, the solution is degenerate, as can be seen
  from Eqn.~\ref{eq:Lnu}.
\item Inverse Compton cooling of CR electrons on CMB photons introduces a scale
  that depends on the CMB-equivalent magnetic field strength,
  $B_\rmn{CMB}\simeq 3.2\,\mu$G; if magnetic field values are close to this 
  value, the resulting synchrotron emission depends non-linearly on the inferred
  magnetic field strengths. Hence, we iterate, repeating the previous steps
  until convergence for the minimum magnetic field $B_0$.
\end{enumerate}

\begin{table}[t]
\begin{center}
  \caption{\label{tab:B} Constraints on magnetic fields in the hadronic model.}
\begin{tabular}{ccccc}
\hline\hline
\phantom{\Big|}
           & \multicolumn{4}{c}{Minimum magnetic field, $B_{0,\rmn{min}}~[\mu\rmn{G}]$:} \\
$\alpha_B$ & \multicolumn{4}{c}{$\alpha$}\\
           & \quad 2.1 \quad & \quad 2.2 \quad & \quad 2.3 \quad & \quad 2.5 \quad \\
\hline\\[-0.5em]
0.3 & ~~5.86 & 4.09 & 3.15 & 2.06 \\ 
0.5 & ~~8.62 & 6.02 & 4.63 & 3.05 \\
0.7 & 13.1~~ & 9.16 & 7.08 & 4.68 \\[0.25em]
\hline
\phantom{\Big|}
           & \multicolumn{4}{c}{Corresponding $X_{\CR}\, (100\,\rmn{kpc})~[\%]$:} \\
\hline\\[-0.5em]
0.3 & 1.7 & 2.5 & 4.9 & 26.7 \\ 
0.5 & 1.7 & 2.5 & 4.8 & 26.1 \\
0.7 & 1.6 & 2.3 & 4.5 & 23.6 \\[0.25em]
\hline
\end{tabular}
\end{center}
{\bf Note.} Constraints on magnetic fields in the hadronic model of
    the Perseus radio mini-halo and the corresponding CR-to-thermal pressure ratio (at
    the largest emission radius of 100 kpc) such that the model reproduces the observed
    radio surface brightness profile.
\end{table}

Table~\ref{tab:B} gives the resulting lower limits for $B_0$ that depend
sensitively on the assumptions of $\alpha$ and $\alpha_B$.   Its behavior 
can be understood as follows. (1) The hardest CR spectral indices
correspond to the tightest limits on $B_0$. This is because for an UL for 
CR energies of around 8 TeV (as probed by 1~TeV gamma-rays) and a CR population with a
softer spectral index (larger $\alpha$) there is a comparably larger fraction of
CRs at 25 GeV available which produce more radio-emitting electrons. Thus, 
lower magnetic field strengths can be accommodated while still matching the observed
synchrotron flux. (2) For a steeper magnetic decline (larger $\alpha_{B}$), the
CR number density must be higher to match the observed radio emission
profiles.  This would yield a higher gamma-ray flux so the ULs are
more constraining. This implies tighter lower limits for $B_{0}$.

The inferred values for the minimum magnetic field strengths in
Table~\ref{tab:B} range from 2--13~$\mu$G for the values of $\alpha$ and
$\alpha_B$ used in this study and suggested by radio observations. These are
much lower than the thermal equipartition value in the center of Perseus,
$B_\rmn{eq,0}\simeq 80\,\mu$G or magnetic field estimates from the Faraday RM
values on small scales \citep{2006MNRAS.368.1500T}. In Table~\ref{tab:B}, we
additionally give the corresponding values for the CR-to-thermal pressure ratio (at the
largest emission radius at 100~kpc) such that the model reproduces the observed
radio surface brightness profile.\footnote{Note that in this section, radial
  profiles of $X_\CR$ are uniquely determined by the adopted model for the
  magnetic field and the observed synchrotron surface brightness profile of the
  radio mini-halo.  This differs from the simplified analytical CR model where
  $X_\CR=\rmn{const.}$ (Sect.\ref{sec:simplified}) and contrasts with the
  simulation-based model where $X_\CR(r)$ is derived from cosmological cluster
  simulations (Sect.\ref{sec:sim}). For most of the cases studied here, the
    CR-to-thermal pressure ratio profiles vary within a factor of two, and flat profiles
    of both the relative magnetic and CR pressure profiles are able to reproduce
    the observed radio surface brightness profile \citep[see Fig. 2
    of][]{2004MNRAS.352...76P}.} Since they are derived from flux upper
limits they are also ULs on the CR-to-thermal pressure ratio.  The corresponding values
for $X_\CR$ in the cluster center are lower than 5\% for the entire parameter
space probed in this study. We conclude that there is still considerable leeway
for the hadronic model as an explanation of the radio mini-halo emission.

\section{Conclusions}
\label{sec:conclusions}

MAGIC observed the Perseus cluster for a total of $\sim85$~hr of high quality
data between October 2009 and February 2011. This campaign resulted in the
detection of the head-tail radio galaxy IC~310 \citep{2010ApJ...723L.207A} and
the detection of the central radio galaxy NGC~1275 \citep{2011arXiv1112.3917M}. 
No significant excess of gamma-rays was detected from the cluster
central region at energies above 630~GeV where the NGC~1275 emission vanishes.  
The flux UL for the CR-induced emission above 1~TeV, for a region of radius of 
$0.15^\circ$ around the cluster center, corresponds to
$1.38 \times10^{-13}$~cm$^{-2}$~s$^{-1}$.

We constrain the CR population and the magnetic field strength in Perseus using
contrasting models that differ in their assumptions and are constructed to
circumvent our lack of knowledge of the true distributions. (1) We use a
simplified ``isobaric CR model'' with constant CR-to-thermal pressure fraction
and power-law momentum spectrum --- a model that has been widely used in the
literature.  (2) This is complemented by an analytical approach based on
cosmological hydrodynamical simulations \citep{2010MNRAS.409..449P} that provide
a CR distribution, combined with the observed density profile. (3) We finally
use a pragmatic approach that models the CR and magnetic field distributions to
reproduce the observed emission profile of the Perseus radio mini-halo.

Within the simplified analytical approach we can constrain the CR-to-thermal
pressure, $X_\CR< 0.8\%$ and 12\% (for a CR or gamma-ray spectral index,
$\alpha$, varying between 2.1 and 2.5). For the spectral index at TeV energies
of $\alpha=2.2$, favored by simulations, we find that $X_\CR<1.1\%$. The
simulation-based approach gives $X_\CR<1.7\%$.  This latter value is a factor of
1.5 less constraining because of the concave curvature of the simulated spectrum
that has higher partial pressure toward GeV energies relative to a pure
power-law spectrum. This constraint is a factor of 1.25 below the simulation
model and -- for the first time -- limiting the underlying physics of the
simulation.  This could either indicate that the maximum acceleration efficiency
of CRs relative to the total dissipated energy at strong structure formation
shocks is $<50\%$ (i.e., smaller than the value assumed in the simulations) or
may point to CR streaming and diffusion out of the cluster core region that
lowers the central $X_\CR$ values \citep{2011A&A...527A..99E}. The observed
spiral X-ray structure in the central cluster region suggests that Perseus is
currently in a relaxation state following a past merging event. If a net outward
CR transport is indeed correlated with a dynamical relaxation state of a
cluster, this would render CR transport a plausible agent that lowers the
gamma-ray flux in comparison to the simulation model that neglects this
mechanism.

Our third, pragmatic, approach employed two different assumptions. First,
adopting a strong magnetic field everywhere in the radio-emitting region ($B\gg
B_\rmn{CMB}$) yields the minimum gamma-ray flux, $F_{\gamma,\rmn{min}}$, in the
hadronic model of radio mini-halos. We find $F_{\gamma,\rmn{min}}$ to be a
factor of 2 to 18 below the MAGIC ULs for spectral indices varying between 2.1
and 2.5. For $\alpha=2.2$, following the universal CR model, the minimum
gamma-ray flux is a factor of 3.2 lower than the MAGIC ULs.  Second, by matching
the radio emission profile (i.e., fixing the radial CR profile for a given
magnetic field model) and by requiring the pion-decay gamma-ray flux to match
the MAGIC flux ULs (i.e., fixing the normalization of the CR distribution) we
obtain lower limits on the magnetic field distribution under consideration.
Recall that we employ a parametrization of the magnetic field of $B=B_0\,
(n/n_0)^{\alpha_B}$.  The inferred values range from $2~\mu\rmn{G} \leq
B_{0,\rmn{min}} \leq13~\mu$G for the parameter space spanned by the magnetic
field strength radial index, $\alpha_B$, and CR spectral index, $\alpha$. Since
$B_{0,\rmn{min}}$ is smaller than recent field strengths estimates through
Faraday RM studies in cool core clusters \citep[e.g.,][]{2011A&A...529A..13K},
this argues that the hadronic model is an interesting possibility in explaining
the radio mini-halo emission. This displays the potential of future gamma-ray
observations of Perseus to further refine the parameters of the hadronic model
and for eventually assessing its validity in explaining radio (mini-)halos. This
is true for the currently operating Cherenkov telescopes and for the future
planned Cherenkov Telescope Array whose sensitivity is planned to be about an
order of magnitude higher than current instruments.

\begin{acknowledgements}
  We thank the anonymous referee for valuable comments.
  We would like to thank Andrey Kravtsov for the useful comments on the
  paper. We would like to thank the Instituto de Astrof\'{\i}sica de Canarias
  for the excellent working conditions at the Observatorio del Roque de los
  Muchachos in La Palma.  The support of the German BMBF and MPG, the Italian
  INFN, the Swiss National Fund SNF, and the Spanish MICINN is gratefully
  acknowledged. This work was also supported by the Marie Curie program, by the
  CPAN CSD2007-00042 and MultiDark CSD2009-00064 projects of the Spanish
  Consolider-Ingenio 2010 programme, by grant DO02-353 of the Bulgarian NSF, by
  grant 127740 of the Academy of Finland, by the YIP of the Helmholtz
  Gemeinschaft, by the DFG Cluster of Excellence ``Origin and Structure of the
  Universe'', by the DFG Collaborative Research Centers SFB823/C4 and SFB876/C3,
  and by the Polish MNiSzW grant 745/N-HESS-MAGIC/2010/0. C.P. gratefully
  acknowledges financial support of the Klaus Tschira Foundation.
  A.P. acknowledges NSF grant AST 0908480 for support.
\end{acknowledgements}

\bibliographystyle{aa}
\bibliography{bib_file}


\end{document}